\documentclass[prl,twocolumn,showpacs]{revtex4}  
\usepackage{graphicx}  

\begin{document}


\title{Possibility of Observing Spiral Scattering of Relativistic Particles
in a Bent Crystal}

\author{G.\,V.\,Kovalev \/\thanks
{e-mail: kovalevgennady@qwest.net}}



\affiliation{North Saint Paul,  MN 55109, USA }


\date{Dec. 26, 2008}

\begin{abstract}
The peak position, impact-parameter range, and optimal conditions for observing spiral scattering of relativistic particles in a uniformly bent crystal are estimated. The existence of spiral scattering with a square-root singularity is pointed out. In this case, the secondary process of volume capture to the channeling mode is
absent and the conditions for observing this effect are most favorable.
\end{abstract}

\maketitle


 The phenomenon of spiral scattering occurs due to
the appearance of a negative logarithmic singularity of
the classical deflection function $\chi(b)$ of a particle or
light ray for a certain impact parameter  $b=b_s$ \cite{FordWheeler_1959_1} . Resonance scattering is a quantum mechanical analog of
spiral scattering \cite{berry_mount_1972}. However, resonance scattering
includes a wider class of quantum-mechanical phenomena. In particular, it can appear in the scattering of fast particles by a cylindrical well (see, e.g., \cite{kal_kov_79}),
whereas classical spiral scattering by such a potential is
absent\cite{kov08_1}. 
\begin{figure}
	\centering
		\includegraphics{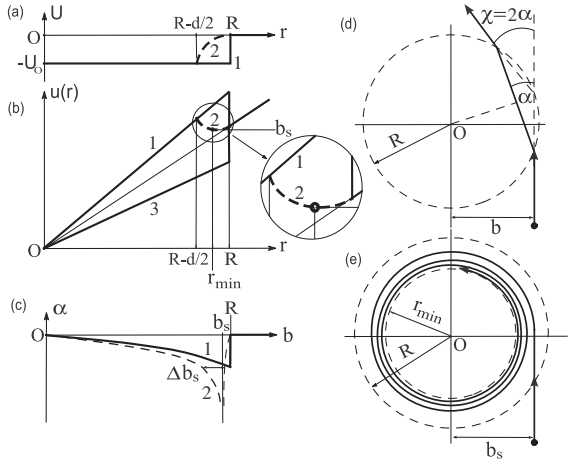}
	\caption{ (a) Square-well potential shown by solid curve 1 and
a smoothed well given by dashed curve 2. (b) The function
$u(r)$ given by Eq. (3) for the (line 1) well, (line 2) smoothed
well, and (line 3) barrier. (c) The deflection function $\alpha=\chi/2$. (d) The trajectory of a particle in the square well. (e) The spiral trajectory of a particle in the smoothed
square well.}
	\label{fig:CentralWells}
\end{figure}

To illustrate this feature and determine the
spiral-scattering boundaries, let us compare scattering
by a cylindrical or spherical potential well of radius $ R$
and depth $-U_0$ and scattering by a well of the same depth
and radius, but with a smoothed parabolic edge of
width $d/2$ ($d << R$) (curves 1 and 2, respectively, in
Fig. 1a):
\begin{eqnarray}
U(r)= - U_0 \frac{4}{d^2}
\left \{\begin{array}{ll}
0,&     \; R < r;\\  
(r-R)^2,&   \; (R-\frac{d}{2}) < r < R;\\ 
\frac{d^2}{4},& \;  0 < r < (R-\frac{d}{2}).
\end{array} \right.
\label{SmoothEdgePotential}
\end{eqnarray} 
The classical deflection function
\begin{eqnarray}
\chi(b)=2\alpha(b)=\pi-2b\int^{\infty}_{r_{o}}\frac{d r }{r \sqrt{r^2[1-\phi(r)]-b^2}},
\label{deflection_function}
\end{eqnarray}
where  $\phi(r)={2U(r) E}/({p_{\infty}^2 c^2})$  and  $b, U(r), E, p_{\infty}$  are
the impact parameter, centrally symmetric potential,
total energy, and momentum of a particle at infinity,
respectively, can be calculated for both cases, but the
absence of spiral scattering for a square well can be
seen directly in the plot (curve 1 in Fig. 1b) of the
function
\begin{eqnarray}
u(r)=r \sqrt{1-\phi(r)},
\label{u_function}
\end{eqnarray}
which has a finite step and a local minimum for $r = R$.
Therefore, the derivative   $u(R)^{'}$ is indeterminate at this
point. Deflection function (2) is expressed in terms of
function (3) as
\begin{eqnarray}
\chi(b)=\pi-2b\int^{\infty}_{r_{o}}\frac{d r }{r \sqrt{u(r)-b}\sqrt{u(r)+b}},
\label{deflection_function1}
\end{eqnarray}
where, as before, the turning point $r_0$ is determined
from the equation $u(r_0) = b$. In the case of the
smoothed potential (curve 2), the function $u(r)$ has a
smooth local minimum at the point rmin, at which
$u'(rmin) = 0$. If the impact parameter coincides with
this minimum, i.e.,
\begin{eqnarray}
b_s = u(r_{min}),
\label{SpiralImpact}
\end{eqnarray}
both the radial velocity and radial acceleration of the
particle become zero (for more detail, see [1, 5]) and
the conditions for spiral scattering are implemented.
Here, the function $u(r)$ in the vicinity of the minimum
$r_{min}$ is represented in the form
\begin{eqnarray}
u(r)\approx u(r_{min}) + \frac{u(r_{min})^{''}}{2}(r-r_{min})^2,
\label{u_functionAppr}
\end{eqnarray}
while the local-minimum point $r_{min}$ determined from
the condition $u'(r_{min}) = 0$ is specified by the equation
\begin{eqnarray}
1-\phi(r_{min})-\frac{r_{min}}{2} \phi(r_{min})^{'} = 0.
\label{gen_r_min_eq}
\end{eqnarray}
This equation holds for any potential; for Eq. (1), it
reduces to the simple form
\begin{eqnarray}
2\hat{r}^2_{min}-3\hat{r}_{min}+1+\delta = 0.
\label{r_min_eq}
\end{eqnarray}
Here,
\begin{eqnarray}
\hat{r}=\frac{r}{R}, \;\; \hat{d}=\frac{d}{R}, \;\;  \delta= \frac{\hat{d}^{2}}{4 |\phi_o|}, \;\; \phi_o=\frac{-2U_0 E}{p_{\infty}^2 c^2}
\label{Notations}
\end{eqnarray}
and $θ_L = \sqrt{|\phi_o|} $
 is the Lindhard channeling angle. In
the case of the potential given by Eq. (1), semi-channeling can occur when a particle moves near the 
potential edge and is successively reflected from its
inner wall. Hereafter, a quantity with the symbol   $\hat{}$ 
denotes the corresponding quantity without this symbol divided by $R$.

Only one, physically meaningful, of the two solutions of Eq. (8) with the plus
sign of the square root should be retained:
\begin{eqnarray}
\hat{r}_{min}=\frac{3}{4} +  \frac{\sqrt{1-8\delta}}{4}.
\label{r_min}
\end{eqnarray}               
Note the important condition that should be in
the range$1- \hat{d}/2 \leq \hat{r}_{min} \leq 1$ (see Fig. 1); otherwise, the
local minimum does not exist. The lower boundary $1- \hat{d}/2 = \hat{r}_{min}$
 substituted into the left-hand part of Eq. (10) yields the critical parameter  $\delta_c$ for the observation  of the spiral scattering of relativistic particles by
a smoothed well:
\begin{eqnarray}
\delta \leq \delta_c= \frac{\hat{d}(1 -\hat{d})}{2}.
\label{criteria_delta}
\end{eqnarray}
Since, according to Eq. (9), $|\phi_o|= \frac{\hat{d}^{2}}{4\delta }$, the corresponding criterion for the squared Lindhard angle
 $|\phi_{o,c}|$ can also be written as
\begin{eqnarray}
|\phi_{o}| \geq |\phi_{o,c}| = \frac{\hat{d}}{2(1 -\hat{d})}.
\label{criteria_Squar_Lindhard}
\end{eqnarray}
Next, since the smooth-edge region is small, $\hat{d} << 1$ (or  $d << R$), Eqs. (11) and (12) reduce to the inequalities
 $\delta \leq  \hat{d}/2$,   $|\phi_{o}| \geq \hat{d}/2$, respectively. It is easy to see
that these constraints are equivalent to the criterion
$R \geq R_c= {p_{\infty}^2c^2 d}/(4U_0 E)=d/(2 \theta^{2}_{L})$. It was noted in [4, 5]
that this criterion is equivalent to the Tsyganov criterion for the existence of the channeling effect in a bent crystal. Since 
$\hat{d} << 1$, $\delta <<  1$ and Eq. (10) can be
approximately written as $\hat{r}_{min} \cong 1-\delta$.

For comparison, the quantity $\alpha(\hat{b})=\chi(\hat{b})/2$  for
the square well in the range $0\leq\hat{b}\leq1$  is given by the
expression (see Section 19 in [6] or Eq. (12) in [5])
\begin{eqnarray}
\alpha(\hat{b}) = \arcsin(\frac{\hat{b}(\sqrt{1-\hat{b}^2}-\sqrt{1-\phi_{o}-\hat{b}^2})}{{\sqrt{1-\phi_{o}}}}).
\label{deflection_Well}
\end{eqnarray}
This corresponds to curve 1 in Fig. 1c. All deflection
angles for the square well are negative with the maximum absolute value
\begin{eqnarray}
|\alpha(1)| = \arcsin(\frac{\sqrt{-\phi_{o}}}{{\sqrt{1-\phi_{o}}}}), 
\label{deflection_WellMIN}
\end{eqnarray}
achieved near the well edge $\hat{b}=1$. Under the condition 
 $\sqrt{|\phi_{o}|}<<1$, this angle is $|\alpha(1)| \approx \theta_{L}$. A typical trajectory of particles in such a potential is shown in
Fig. 1d. In the case of the smoothed well given by
Eq. (1), deflection function (2) is given by a lengthy
formula with elliptical functions. However, if approximation (6) is used, the deflection function for the
impact parameters 
 $\hat{b}_s < \hat{b} < 1$  takes the form
\begin{eqnarray}
\alpha(\hat{b}) =\frac{\pi}{2}-\arcsin(\hat{b})-\sqrt{\frac{\hat{b}}{\hat{u}(\hat{r}_{min})^{''}}}*\nonumber \\ \frac{\ln(\frac{2(\hat{r}_0-\hat{r}_{min})}{(\sqrt{\hat{r}_0(1+\hat{r}_0-2\hat{r}_{min})}-\sqrt{(1-\hat{r}_0)(2 \hat{r}_{min}-\hat{r}_0)})^2})}{\sqrt{\hat{r}_{min}^2-(\hat{r}_0-\hat{r}_{min})^2}},
\label{deflection_EdgeWell1}
\end{eqnarray}
where $\hat{r}_0=\hat{r}_{min} + \sqrt{2(\hat{b}-\hat{u}(\hat{r}_{min}))/\hat{u}(\hat{r}_{min})^{''}}$. The
deflection function for the impact parameters $0 < \hat{b} < \hat{b}_s$  has the different form
\begin{eqnarray}
\alpha(\hat{b}) =\arcsin(\frac{\hat{b}}{A\sqrt{1+\hat{d}^2/(4\delta)}})-  \arcsin(\hat{b})-\nonumber \\
\sqrt{\frac{\hat{b}}{C\hat{u}(\hat{r}_{min})^{''}}} \ln(\frac{2C+AB+2\sqrt{C}\sqrt{A^2+AB+C}}{A(2C+B+2\sqrt{C}\sqrt{1+B+C})}),
\label{deflection_EdgeWell2}
\end{eqnarray}
where $A=1-\hat{d}/2$, $B=-2\hat{r}_{min}$, $C=\hat{r}^2_{min}+2(\hat{u}(\hat{r}_{min})-\hat{b})/(\hat{u}(\hat{r}_{min})^{''})$.

If  $\hat{b} \rightarrow \hat{b}_s=\hat{u}(\hat{r}_{min})$, the deflection angles given
by Eqs. (15) and (16) tend logarithmically to$-\infty$  (see
curve 2 in Fig. 1c); i.e., the particle orbits the potential
center by an angle exceeding maximum angle (14) for
a square well. Under the condition   $\hat{b} = \hat{b}_s$, spiral scattering appears; in this case, the relativistic particle
does not have the outgoing branch and is located near
the potential for an infinitely long time (see Fig. 1e)
approaching the limit cycle $\hat{r} \rightarrow \hat{r}_{min}$.   Only one trajectory satisfies the spiral-scattering criterion. Trajectories with impact parameters close to  $\hat{b}_s$ 
 have the usual ingoing and outgoing branches, but can follow
the circle over an angle exceeding the Lindhard angle
(14).

Let the spiral-scattering range be defined as the
impact-parameter range $\Delta b_s$ near $b = b_s$, in which the
absolute values of negative angles (15) and (16) exceed
maximum angle (14) for the square well (see Fig. 1c).
Such a definition is useful since it distinguishes the
spiral scattering from both refraction ranging within
the angle given by Eq. (14) and volume reflection,
which has a positive sign [7], but does not exceed the
angle given by Eq. (14). The spiral scattering is also
different from channeling (see below). The substitution of Eq. (10) into Eq. (5) yields the exact value for
the spiral impact parameter $\hat{b}_s$
 in the case of potential  (1):
\begin{eqnarray}
\hat{b}_s=\frac{3+\sqrt{1-8\delta}}{4}\sqrt{1+\frac{(1-\sqrt{1-8\delta})^2}{16\delta}} .
\label{SpiralImpactForSmoothEdgePotential}
\end{eqnarray}
If $\delta << 1/8$,  $\hat{b}_s\approx 1- \delta/2$.

The right and left boundaries of the spiral-scattering range are determined as the roots of the transcendental equation
\begin{eqnarray}
\alpha(\hat{b}) = -\arcsin(\frac{\sqrt{|\phi_{o}|}}{{\sqrt{1+|\phi_{o}|}}}),
\label{TransendentalEq}
\end{eqnarray}
with $\alpha(\hat{b})$ given by Eqs. (15) and (16), respectively. By
solving the simpler transcendental equation $c_1 \ln{(\Delta \hat{b}_{sr})} +c_2=0$
 (where c1 and c2 are constants),
which is obtained by retaining only the first terms in
the expansions of all functions in Eq. (15) in the power
series of $(\hat{b}-\hat{b}_s)$  in the right neighborhood of $\hat{b}_s$, the
right part $\Delta \hat{b}_{sr}=\hat{b}_{r}-\hat{b}_{s}$ 
 of the region  $\Delta \hat{b}_s$ is determined in the form
\begin{eqnarray}
\Delta \hat{b}_{sr}=2\hat{u}(\hat{r}_{min})^{''}\delta^2  \exp\biggl
(-\sqrt{\frac{\hat{u}(\hat{r}_{min})^{''}}{\hat{b}_{s}}}*\nonumber \\ \Bigl
(\frac{\hat{d}}{\sqrt{\delta}}
+2  \sqrt{1-\hat{b}_{s}^2}  \Bigr)\biggr ).
\label{RightBoundary}
\end{eqnarray}
Hereafter, it is taken into account the conditions $|\phi_{o}|<<1$ and $\delta<<1/8$, which are certainly valid in the relativistic case.

To calculate the left-hand part  $\Delta \hat{b}_{sl}=\hat{b}_{s}-\hat{b}_{l}$  in the
entire range given by Eq. (12), it is necessary to retain
a few terms in the expansion of Eq. (16) in the power
series of $(\hat{b}-\hat{b}_s)$, since the impact parameter $\hat{b}_s$
approaches the inflection point $\approx \hat{d}/4$ of the function
 $\hat{u}(\hat{r})$ with a variation in the parameter $|\phi_{o}|$. In the case
where they are kept, the solution of the transcendental
equation  $c_1 \ln{(c_3 \Delta \hat{b}_{sl})}+ \Delta \hat{b}_{sl} +c_2=0$, obtained by
retaining only first two terms in the series, has a solution expressed in terms of the Lambert function  $W(x)$,
which is defined through the relation  $W(x) \exp{W(x)}=x$ [8], as
\begin{eqnarray}
\Delta \hat{b}_{sl}= c_1 W(\frac{\exp{(-\frac{c_2}{c_1})}}{c_1 c_3}),
\label{LeftBoundary}
\end{eqnarray}
where the parameters  $c_1, c_2$ and $c_3$  are
\begin{eqnarray}
c_1=\frac{\sqrt{A^2(1+\phi_0)-\hat{b}^2_{s}}\sqrt{1-\hat{b}^2_{s}}}{(\sqrt{A^2(1+\phi_0)-\hat{b}^2_{s}}-\sqrt{1-\hat{b}^2_{s}})\sqrt{\hat{u}(\hat{r}_{min})^{''}}}  \nonumber \\ 
c_2 =-\sqrt{A^2(1+\phi_0)-\hat{b}^2_{s}}\sqrt{1-\hat{b}^2_{s}} +\nonumber \\
+ \frac{\hat{d}\sqrt{A^2(1+\phi_0)-\hat{b}^2_{s}}\sqrt{1-\hat{b}^2_{s}}}{2\sqrt{\delta (1+\phi_0)}(\sqrt{A^2(1+\phi_0)-\hat{b}^2_{s}}-\sqrt{1-\hat{b}^2_{s}})}\nonumber \\
c_3 =\frac{A}{2 \delta(\hat{d}/2-\delta)\hat{u}(\hat{r}_{min})^{''}}.
\label{LeftBoundaryParam}
\end{eqnarray}
The width of the impact-parameter region $\Delta \hat{b}_{s}$
 where
spiral scattering is significant is determined by the sum
of Eqs. (19) and (20):
\begin{eqnarray}
\Delta \hat{b}_{s}=\Delta \hat{b}_{sr}+\Delta \hat{b}_{sl}.
\label{Boundary}
\end{eqnarray}
The second derivative $\hat{u}(\hat{r}_{min})^{''}$ in Eqs. (19) and (20) is
expressed in terms of $\delta$ as
\begin{eqnarray}
\hat{u}(\hat{r}_{min})^{''}=\frac{4\hat{r}_{min}-3}{\delta \sqrt{1+\frac{(\hat{r}_{min}-1)^2}{\delta}}}=\frac{\sqrt{1-8\delta}}{\delta \sqrt{1+\frac{(1-\sqrt{1-8\delta})^2}{16\delta}}},
\label{2ndDerivative}
\end{eqnarray}
and  $\hat{u}(\hat{r}_{min})^{''} \approx \delta^{-1}$ for small  $\delta$ values.

\begin{figure}[tbp]
	\centering
		\includegraphics{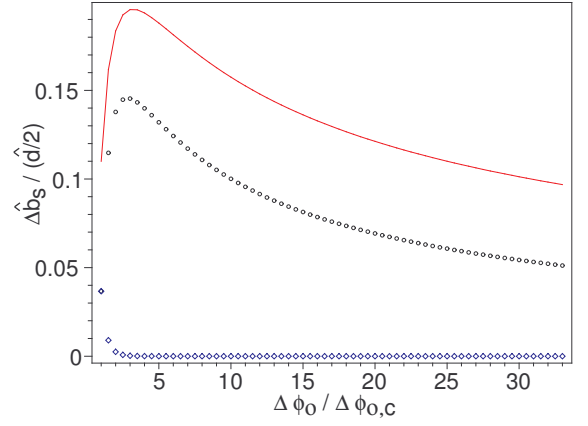}
	\caption{Dimensionless spiral-scattering width  $\Delta \hat{b}_{s}$ versus
the parameter $|{\phi_{o}}|/|{\phi_{o,c}|}$.  The solid curve is the numerical
calculation by  \ref{TransendentalEq}. The circles show the contribution
from $\Delta \hat{b}_{s}$ to $\Delta \hat{b}_{sl}$, according to  \ref{LeftBoundary}. The diamonds show the contribution from $\Delta \hat{b}_{s}$ to $\Delta \hat{b}_{sr}$, according to  \ref{RightBoundary}.}
	\label{fig:Width}
\end{figure}

In addition to approximate formulas (19) and (20)
for the right and left boundaries of the spiral-scattering
region, respectively, transcendental equations (18)
were also solved numerically. The result is shown by
the solid curve in Fig. 2. The numerical solution for
the right-hand boundary (diamonds) coincides with
high accuracy with Eq. (19). Formula (20) for the left-hand boundary (circles) provides an underestimated
result, but the behavior of the curve is the same as predicted by the accurate numerical solution. In addition,
the contribution from the right-hand part of the spiral-scattering range to the total width is much smaller than the contribution from the left-hand part, 
except for
the vicinity of the critical point $|\phi_{o}|=|\phi_{o,c}|$. 
As a result,
the logarithmic branches of the deflection function on
the right and left sides of  $b = b_s$ are not symmetric for
the considered potential (curves 2 and 3 in Fig. 3).

\begin{figure}[tbp]
	\centering
		\includegraphics{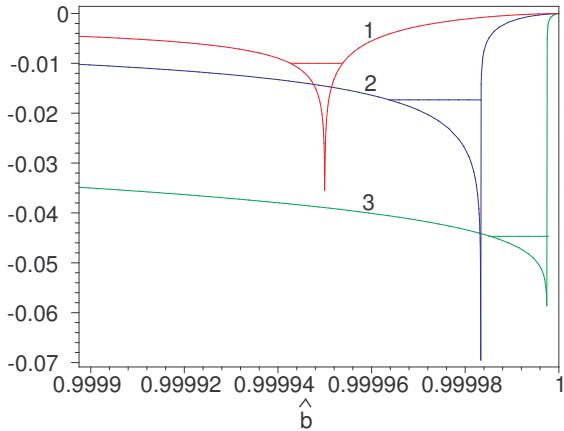}
	\caption{Deflection function $\alpha (\hat{b})$ for the potential given by \ref{SmoothEdgePotential}: 1 - $\hat{d}=2\cdot10^{-4}$, $|\phi_{o}|=10^{-4}$, 2   - $\hat{d}=2\cdot10^{-4}$, $|\phi_{o}|=3\cdot10^{-4}$,  3 - $\hat{d}=2\cdot10^{-4}$, $|\phi_{o}|=2\cdot10^{-3}$. The horizontal line segments for each curve correspond to the refraction by the angle $\theta_{L}=\sqrt{|\phi_{o}|}$.  The
widths of these segments correspond to the spiral-scattering and deflection boundaries.}
	\label{fig:Deflection}
\end{figure}

Note an important feature of the width of the spiral-scattering region. Let the potential parameters $R$,
$U_0$,  and $d$ be fixed and the relativistic particle energy
decrease so that $|\phi_{o}|$ increases from critical value (12)
to a few tens of $|\phi_{o,c}|$ without violating the condition
$|\phi_{o}|<<1$.  Then, the width  $\Delta \hat{b}_{s}$
 of the spiral-scattering
region increases rapidly up to the maximum for $|\phi_{o}|\approx 3|\phi_{o,c}|$ and then decreases to zero. In this specific case,
the normalized maximum value$\Delta \hat{b}_{s,max}/(\hat{d}/2)$ is
$\approx 19 \%$ of the width $\hat{d}/2$  (see Fig. 2) and its distance
from the critical energy is  $|\phi_{o}|/|\phi_{o,c}|$. Thus, the position of the maximum on this scale is independent of
the particle energy. Moreover, the normalized width
\begin{eqnarray}
\Delta \hat{b}_{s}/(\hat{d}/2)=f(|\phi_{o}|/|\phi_{o,c}|)
\label{InvariantEq}
\end{eqnarray}
is a universal function determined only by the shape of
the potential. At the same time, the value of the
parameter $|\phi_{o}|$ for which the maximum is reached evidently depends on the particle energy. However, the
above-mentioned scaling invariance holds in this case.
This feature is seen in Fig. 4, which presents the width
of the spiral-scattering range for the ring potential
given by Eq. (1) with the hollow core ($U(r)=0$ for $r< R-d/2$). This potential ensures maximum volume
reflection in comparison with other ring potentials of
the depth $–U_0$ and width $d$ constructed of two inversed
parabolas. Thus, this potential provides a minimum
estimate for the possible width of spiral scattering for
an actual crystal. The value $\Delta \hat{b}_{s,max}/(\hat{d}/2)\approx 7 \%$ and
position $|\phi_{o}|/|\phi_{o,c}|\approx 1.3$  of the maximum of the spiral
scattering width are different in this case, but the scaling invariance holds.

\begin{figure}[tbp]
	\centering
		\includegraphics{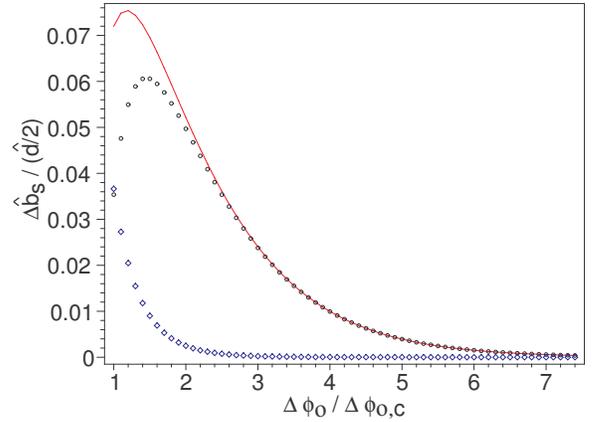}
	\caption{Solid curve is the dimensionless spiral-scattering
width   $\Delta \hat{b}_{s}/(\hat{d}/2)$ calculated numerically for the ring
potential with a half-parabola by \ref{TransendentalEq}.  The circles
show the contribution from $\Delta \hat{b}_{s}$ to $\Delta \hat{b}_{sl}$.The diamonds
show the contribution from $\Delta \hat{b}_{s}$ to $\Delta \hat{b}_{sr}$.}
	\label{fig:Width1}
\end{figure}

Note that the $\alpha (\hat{b})$ plots for $\hat{d}$ and $|\phi_{o}|$  values other
than those shown in Fig. 3 have the same form if $\alpha$ and $\hat{b}$
 are normalized to  $\theta_{L}$ and $\hat{d}/2$, respectively, due to
the above mentioned scaling invariance. It is seen that
the width of the spiral-scattering range for large $|\phi_{o}|$ 
values ($|\phi_{o}|=20 \cdot|\phi_{o,c}|$, curve 3 for the relatively low
energies) decreases and tends to zero in the limit. In
this case, the deflection function tends to Eq. (13) for
the square-well scattering, for which the smoothness
of the edge is not important and the deflection angles
$\alpha (\hat{b})$ are no larger than  $\theta_{L}$ (the angle $\chi (\hat{b})$ is no larger
than $2 \theta_{L}$). The width of the spiral-scattering range,
$\Delta \hat{b}_s$, increases with energy (with decreasing $|\phi_{o}|$). This
width on curve 2 is maximal for $|\phi_{o}|=3|\phi_{o,c}|$. Then, the
width $\Delta \hat{b}_s$  decreases. Curve 1 corresponds to the critical energy  $|\phi_{o}|=|\phi_{o,c}|=\hat{d}/2$, for which channeling is already impossible since the effective well is absent.
Meanwhile, spiral scattering exists for this energy.

\begin{figure}[tbp]
	\centering
		\includegraphics{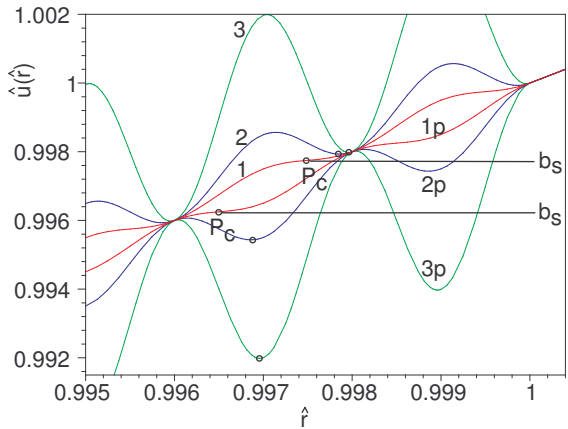}
	\caption{Function $\hat{u}(\hat{r})=\hat{r}\sqrt{1\pm\frac{\phi_0 (1-\cos(2 \pi (1-\hat{r})/\hat{d}))}{2}}$ for a crystal potential for various particle energies and fixed
parameters  $R$, $U_0$, $d$.  $1$, $2$, $3$ - negatively charged particles with ($\hat{d}=0.002$,    $\phi_0=0.001,\; 0.003,\; 0.01$); $1p$, $2p$, $3p$ - positively charged particles (the minus sign under
the square root in $\hat{u}(\hat{r})$ );  $P_c$ are the inflection points of
plots $1$ and $1p$ for the critical energies corresponding to the
Tsyganov criterion.}  
	\label{fig:InflactionPoint}
\end{figure}

The last feature is due to the existence of a singularity of the deflection function for any smooth periodic
potential similar to that shown in Fig. 5 if the critical
conditions corresponding to the equality in Eqs. (11)
and (12) are exactly satisfied. However, in contrast to
the logarithmic singularity considered by Ford and
Wheeler [1] and given by expansion (6), the singularity
type changes. In this case, an inflection point ( $P_c$ in
Fig. 5) appears instead of the local minimum; i.e., the
second derivative $u^{''}(r_{min})$ is also zero at this point and
the series expansion of $u(r)$  in the vicinity of the minimum $r =r_{min}$  becomes
\begin{eqnarray}
u(r)\approx u(r_{min}) + \frac{u(r_{min})^{'''}}{6}(r-r_{min})^3.
\label{u_functionAppr2}
\end{eqnarray}
It is easy to see that integral (4) for $b=u(r_{min}) $ has the
inverse square-root singularity
\begin{eqnarray}
\chi(b)  \approx  -\frac{1}{\sqrt{|u(r_{min})-b|}},
\label{deflection_function2}
\end{eqnarray}
Hence, the spiral scattering and deflection should take
place for $|\phi_{o}|=|\phi_{o,c}|$.

Note that although the effective
force is zero at the local minimum and inflection
points, the spiral scattering and deflection by a potential without the centrifugal term always take place at
the internal slope of the potential responsible for
attraction (see the local minima in Fig. 5).

These features hold also for a real crystal potential
having one local minimum or inflection point per
period under conditions (11) and (12). In this case, the
position and maximum width are certainly different
and, in addition, the width should be normalized to
the crystal period  $\hat{d}$. This yields the lower estimate for
the maximum width $\Delta \hat{b}_s/\hat{d}\approx 3.5 \%$  (ring potential)
and the upper estimate $\Delta \hat{b}_s/\hat{d}\approx 9.5 \%$ (potential with
edge). However, the scaling invariance should also
hold for these potentials.

For experiments, this means that if, e.g., the crystal curvature radius $R$ is changed,
then the particle energy $E$ for a given crystal potential
can always be chosen so that the parameter $|\phi_{o}|$ satisfies
the criterion for which the spiral-scattering width is
maximal. At the same time, the bending radius $R$ maximizing the spiral-scattering width can always be pointed out for any relativistic particle energy $E$.

It should be noted that the multiple scattering and
dissipation processes in a real crystal do not eliminate
the singularity associated with spiral scattering. This
was demonstrated for nuclear reactions such as, e.g.,
$^{40}Ar$ + $^{232}Th$, in which deep inelastic processes were
interpreted in terms of negative spiral-scattering
angles with allowance for friction forces [9]. This is
explained by the fact that friction forces deform particle trajectories, but the spiral-scattering range  $\Delta \hat{b}_s$ remains unchanged.

Owing to the absence of thermal vibrations and the
low electron density in the region of local minima, the
volume capture of negatively charged particles is much
smaller and the spiral-scattering width is much greater
than the respective quantities for positively charged
particles. Hence, negatively charged particles
deflected along the crystal bending by an angle larger
than $2 \theta_{L}$ propagate in the spiral-scattering mode. The
above estimates show that the fraction of negatively
charged particles for the maximum spiral-scattering
width is  $3\%$ äî $9\%$. The preliminary experiments [10, 11]
on the scattering of 180-GeV electrons in $Si$ show that
a fairly large number of particles move at an angle
larger than $>2 \theta_{L}=32$ µrad and do not have the peak
typical of volume capture at the end of the angular size
of the crystal.

The local minima in the case of positively charged
particles moving in weakly bent crystals are in the
region of high electron densities and thermal fluctuations of potentials. Thus, the trapping of spiral particles to the channeling mode seems to be the most
probable secondary process. However, the spiral scattering is also significant in this case since the motion
along the crystal bending ensures the intense volume
capture of the particles in the region  $\Delta \hat{b}_s$. 
For the crystal bending equal to the critical radius $R_c$ when the
channeled states are absent, the presence of particles
with a rotation angle exceeding $2\theta_{L}$ can be undoubtedly interpreted as the presence of spiral scattering in
this case and in the case of negatively charged particles.

I am grateful to Andrea Mazzolari for attracting my
attention to work [10].

\bibliographystyle{jetpl}

\end{document}